\documentclass[useAMS,usenatbib,referee]{mn2e}

\usepackage{graphicx}

\newcommand{\um}{$\mu$m}
\newcommand{\kms}{km s$^{-1}$}
\newcommand{\cmdue}{cm$^{-2}$}
\newcommand{\cmtre}{cm$^{-3}$}

\newcommand{\msun}{M$_{\odot}$}
\newcommand{\ndueh}{N$_2$H$^+$\,}
\newcommand{\hctren}{HC$_3$N\,}
\newcommand{\chtreoh}{CH$_3$OH\,}

\def \arcsec{\hbox{$^{\prime\prime}$}}
\def \arcmin{\hbox{$^{\prime}$}}
\def \degr{\hbox{$^{\circ}$}}

\title[Multiline spectral imaging of dense cores in Lupus]{Multi-line spectral imaging of dense cores in the Lupus molecular cloud}
\author[M. Benedettini et al.]{M. Benedettini$^{1}$\thanks{E-mail: milena@ifsi-roma.inaf.it}, S. Pezzuto$^{1}$, M. G. Burton$^{2}$, S. Viti$^{3}$, S. Molinari$^{1}$, P. Caselli $^{4}$, L. Testi$^{5,6}$\\
$^{1}$INAF -- Istituto di Fisica dello Spazio Interplanetario, Area di Ricerca di Tor Vergata, via Fosso del Cavaliere 100, 00133 Roma, Italy \\
$^{2}$School of Physics, University of New South Wales, Sydney NSW 2052, Australia\\
$^{3}$Department of Physics and Astronomy, University College London, Grower Street, London WC1E 6BT, UK\\
$^{4}$School of Physics and Astronomy, University of Leeds, LS2 9JT Leeds, UK\\
$^{5}$INAF -- Osservatorio Astrofisico di Arcetri, Largo E. Fermi 5, 50125 Firenze, Italy\\
$^{6}$ESO, Karl Schwarschild Strasse 2, 85748 Garching bei München, Germany
}
\begin{document}

\date{Accepted 2011 August 24.  Received 2011 August 23; in original form 2011 February 17}

\pagerange{\pageref{firstpage}--\pageref{lastpage}} \pubyear{2009}

\maketitle

\label{firstpage}

\begin{abstract}

The molecular clouds Lupus 1, 3 and 4 were mapped with the Mopra telescope at 3 and 12 mm. Emission lines from high density molecular tracers were detected, i.e.  NH$_3$ (1,1), NH$_3$ (2,2), N$_2$H$^+$ (1-0), HC$_3$N (3-2), HC$_3$N (10-9), CS (2-1), CH$_3$OH (2$_0$-1$_0$)A$^+$ and CH$_3$OH (2$_{-1}$-1$_{-1}$)E. Velocity gradients of more than 1 \kms\, are present in Lupus 1 and 3 and multiple gas components are present in these clouds along some lines of sight. Lupus 1 is the cloud richest in high density cores, 8 cores were detected in it, 5 cores were detected in Lupus 3 and only 2 in Lupus 4. The intensity of the three species \hctren, NH$_3$ and \ndueh changes significantly in the various cores: cores that are brighter in \hctren\, are fainter or undetected in NH$_3$ and \ndueh and vice versa. We found that the column density ratios HC$_3$N/\ndueh and HC$_3$N/NH$_3$ change by one order of magnitude between the cores, indicating that also the chemical abundance of these species is different. The time dependent chemical code that we used to model our cores shows that the HC$_3$N/\ndueh and HC$_3$N/NH$_3$ ratios decrease with time therefore the observed column density of these species can be used as an indicator of the chemical evolution of dense cores. On this base we classified 5 out of 8 cores in Lupus 1 and 1 out of 5 cores in Lupus 3 as very young protostars or prestellar cores. Comparing the millimetre cores population with the population of the more evolved young stellar objects identified in the Spitzer surveys, we conclude that in Lupus 3 the bulk of the star formation activity has already passed and only a moderate number of stars are still forming. On the contrary, in Lupus 1 star formation is on-going and several dense cores are still in the pre--/proto--stellar phase. Lupus 4 is at an intermediate stage, with a smaller number of individual objects.

\end{abstract}

\begin{keywords}
ISM: molecules -- ISM: abundances -- radio lines: ISM.
\end{keywords}

\section{Introduction}

One of the critical open question in star formation is the accurate determination of the stellar Initial Mass Function (IMF), especially in the low-mass regime, in order to understand its origin and particularly how it is related to the mass distribution of the dense cores where stars form, i.e. the Core Mass Function (CMF). 
This fundamental question is investigated by the means of surveying dense condensations in molecular clouds.
One of the classical tool for detecting dense cores in star forming regions is the search for dust condensations using continuum measurement in the millimetre range (e.g. \citealt{testi98}; \citealt{Johnstone00}; \citealt{motte01}). Alternatively, one can use spectroscopic surveys of dense gas molecular tracers.
Ammonia is one of the best molecules for studying the cool, dense molecular cores where stars form (e.g. \citealt{myers83}; \citealt{benson89}). 
High-density condensations are also mapped in other molecular tracers such as  N$_2$H$^+$ and CS. In particular, N$_2$H$^+$ is known to be a good tracer of the dense centre of the cores, while CS is depleted from the gas phase in the very centre of prestellar cores and preferentially samples the core edge (\citealt{caselli02b}; \citealt{tafalla02}). CS is also a good tracer of extended high density gas and is useful to probe the kinematics of the gas (\citealt{testi00}; \citealt{olmi02}).

The poorly studied Lupus molecular cloud is an interesting target for investigating the low mass star formation process because its star formation regime, in terms of star formation rate and stellar clustering, represents an intermediate case between the heavily clustered sites such as Serpens and Ophiuchus and the more isolated and quiescent sites such as Taurus. Because of its location in the Southern hemisphere (declination from -33\degr to -43\degr), this extended molecular cloud has been less investigated with respect to more famous Northern regions.

The distance of the Lupus star forming region is still subject of debate even if it is clear that it is one of the nearest star forming regions.The most recent distance measurement is from \citet{lombardi08} that estimated a distance of (155$\pm$8) pc. \cite{comeron08} reviewed all the works about the distance determination for the Lupus complex concluding that it has a depth of the same order as its angular extent on the plane of the sky, with varying distances of the different individual structures in the 140 to 200 pc range. He concluded that a a distance of 150 pc is adequate for Lupus 1 and 4 while a value of 200 pc is more appropriate for Lupus 3.

Up to now only a few surveys at poor spatial resolution of a few arcminutes have been carried out in the region in order to study the dense molecular gas distribution (e.g. \citealt{hara99}; \citealt{vilas00}; \citealt{tachihara01}; \citealt{tothill09}). Maps in the (J=1-0) transition of $^{12}$CO and its isotopologues $^{13}$CO and C$^{18}$O have shown that the extended Lupus complex is actually split into nine subgroups. Evidences of on-going star formation have been found in three subgroups, namely Lupus 1, 3 and 4. Lupus 1, with a mass of $\sim$1200~\msun, is the most massive subgroup. About ten C$^{18}$O cores have been identified in Lupus 1 with column densities N(C$^{18}$O)=(5--10)$\times$10$^{14}$ \cmdue\, \cite[]{hara99} that indicate potential sites of star formation. 
Lupus 3 has a mass of about 300 \msun\, and it hosts a rich cluster of T-Tauri stars. \citet{tachihara07} have mapped the cloud in H$^{13}$CO$^+$, showing that no more star formation is expected at the west edge where the T association is located, whereas there are potential sites of star formation at the eastern edge where H$^{13}$CO$^+$ emission has been detected. 
Lupus 4 is the third cloud of the complex that shows evidence of star formation activity, hosting nine C$^{18}$O dense cores with column densities N(C$^{18}$O)=(4--10)$\times$10$^{14}$ \cmdue\, and three H$^{13}$CO$^+$ cores \cite[]{hara99}.
The Lupus 1, 3 and 4 clouds have been mapped with IRAC and MIPS on board Spitzer as part of the ``From molecular clouds to planet-forming disks'' (c2d) Legacy Program (\cite{merin08} for the IRAC data and \cite{chapman07} for the MIPS data). These infrared surveys allowed the identification of the population of Young Stellar Objects (YSOs) in the clouds. Adding also Pre Main Sequence (PMS) objects previously known from others studies, \cite{merin08} found that the total number of  PMS is 17, 124 and 18 in Lupus 1,  3 and 4, respectively and that the Star Formation Rate (SFR) is 4.3, 31.0 and 4.5 \msun Myrs$^{-1}$ in Lupus 1, 3 and 4, respectively, indicating that Lupus 3 has a higher star formation activity then Lupus 1 and 4. 

The Lupus clouds are included in the extended surveys of nearby molecular clouds that are being carried out with both ground and space facilities, namely: 70, 100, 160, 250, 350 and 500 \um\, Gould Belt Herschel survey \cite[]{andre10}; JCMT Gould's Belt Legacy Survey with SCUBA2 and HARP-B (\citealt{atchell05}; \citealt{johnstone04}).

In order to identify the population of pre- and proto-stellar cores, we carried out a molecular survey of the three Lupus subgroups where there is evidence of star formation activity, i.e. Lupus 1, 3 and 4, at millimetre wavelengths in several molecular species that are good tracers of dense gas. The millimetre data are complementary to the other surveys of the region and will facilitate their interpretation. The complete set of data will allow to understand the star formation activity in the Lupus region and more in general the low-mass star formation process from the cores condensation to the protostellar phase.

\section[]{Observations and Data Reduction}
\begin{table}
 \centering
  \caption{List of of the transitions selected for the observation. The transitions that have been detected are in boldface.}
  \begin{tabular}{@{}lcc@{}}
  \hline
  species   & transition    &  $\nu$ (GHz)   \\
  \hline
{\bf HC$_3$N}    &  ({\bf10-9)}	   & {\bf90.979 }\\
CH$_3$CN   &  (5-4)        & 91.971 \\
$^{13}$CS  &  (2-1)        & 92.494 \\
{\bf N$_2$H$^+$} &   {\bf (1-0)}       & {\bf 93.174} \\
$^{13}$CH$_3$OH& (2$_{-1}$-1$_{-1}$)E& 94.405 \\
CH$_3$OH   & (8$_0$-7$_1$)A$^+$&  95.170 \\
CH$_3$OH   & (2$_1$-1$_1$)A$^+$&  95.914 \\
C$^{34}$S  & (2-1)         & 96.412 \\
{\bf CH$_3$OH}   & {\bf (2$_{-1}$-1$_{-1}$)E}& {\bf 96.739} \\
{\bf CH$_3$OH}   & {\bf (2$_0$-1$_0$)A$^+$} & {\bf 96.741} \\
CH$_3$OH   & (2$_0$-1$_0$)E &  96.745 \\
CH$_3$OH   & (2$_1$-1$_1$)E &  96.755 \\
OCS        & (8-7)         & 97.301  \\
CH$_3$OH   & (2$_1$-1$_1$)A$^-$&  97.583 \\
{\bf CS}         & {\bf (2-1)}         & {\bf 97.981}  \\
\hline
CH$_3$OH &  (2$_1$-3$_0$)E &  19.967\\ 
H        &  68$\alpha$     &  20.462 \\
H        &  67$\alpha$     &  21.385 \\
NH$_3$   &  (3,2)	   &  22.834 \\ 
CH$_3$OH &(9$_2$-10$_1$)A$^+$& 23.121 \\ 
H	 &   65$\alpha$	   &  23.404\\
{\bf NH$_3$}&   {\bf (1,1)}&  {\bf 23.694}\\ 
{\bf NH$_3$}&   {\bf (2,2)}&  {\bf 23.722}\\ 
NH$_3$   &   (3,3)	   &  23.870\\
CH$_3$OH & (5$_2$-5$_1$)E  &  24.959\\ 
H        &  63$\alpha$     &  25.686 \\
CCS      &  (2$_2$-1$_1$)  &  25.911\\ 
H        &  62$\alpha$     &  26.939\\
{\bf HC$_3$N}&  {\bf (3-2)}&  {\bf 27.294}\\
\hline
\end{tabular}
\label{obs}
\end{table}

Molecular line surveys of the Lupus 1 and 3 molecular clouds at 3 and 12 mm were carried out with the Mopra telescope. Moreover, the Lupus 4 cloud were also mapped at 12 mm. The observations were executed in two periods: from 17 to 19 July 2008 and from 20 to 26 October 2008.

The observations were carried out in the On The Fly observing mode with the narrow band mode of the UNSW-Mopra Spectrometer(UNSW-MOPS) digital filterbank back-end, and the Monolithic Microwave Integrated Circuit (MMIC) 77 to 116 GHz receiver. UNSW-MOPS has a 8-GHz bandwidth with four overlapping 2.2-GHz subbands, each subband having four dual-polarization 137.5-MHz-wide windows giving a total of sixteen dual-polarization windows. Each window has 4096 channels providing a velocity resolution of 0.11 \kms\, at 94 GHz and 0.41 \kms\, at 22 GHz. We selected the 16 zoom bands in the range between 19.5 and 27.5 GHz and at 12 mm and between 90 and 98 GHz at 3 mm. The selected frequencies are listed in Table \ref{obs}. 

\begin{figure*}
\includegraphics[width=17cm,angle=0]{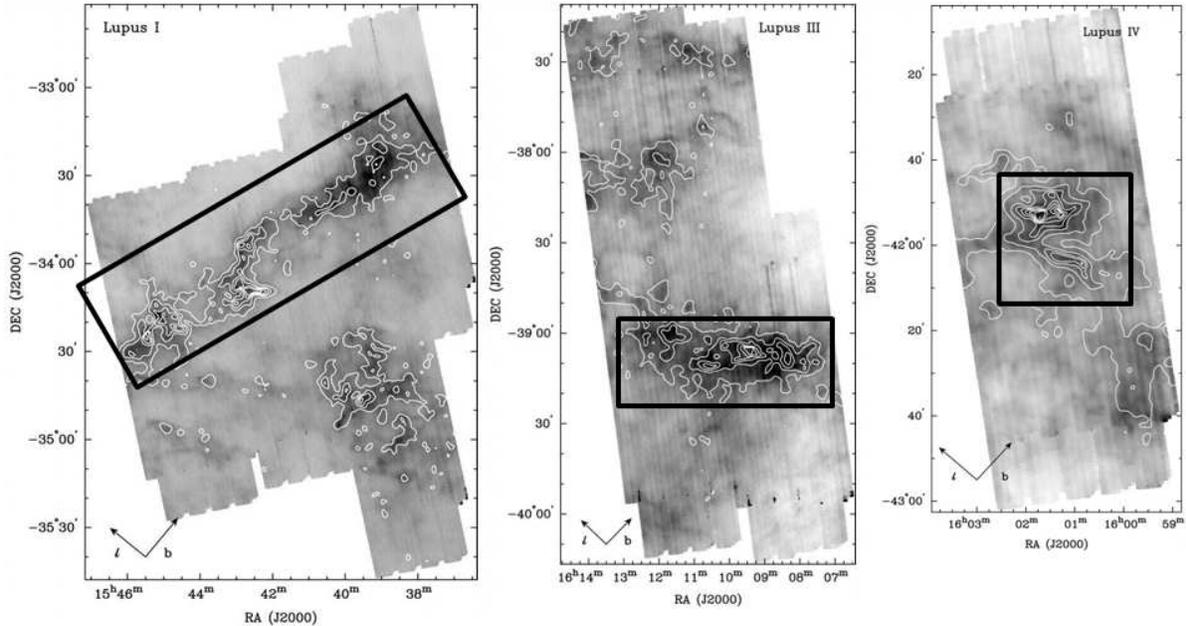}
\caption{Visual extinction map (contours) of the Lupus 1 (left panel), Lupus 3 (middle panel) and Lupus 4 (right panel) molecular clouds overlaid on the Spitzer-MIPS 160 $\mu$m maps \citep{chapman07}. The contours range from 3 mag to 15, 18 and 27 mag for Lupus 1, 3 and 4 respectively; contours steps are at 3 mag. The rectangles indicate the mapped regions.}
\label{maps}
\end{figure*}

Since the beam of the telescope at 12 mm is 2.5\arcmin, at this frequency we were able to map all the zones of the Lupus 1, 3 and 4 clouds with visual extinction larger than 3 mag (see Fig. \ref{maps}), i.e. a region of  130\arcmin$\times$40\arcmin\, in Lupus 1, 70\arcmin$\times$20\arcmin\, in Lupus 3 and 30\arcmin$\times$30\arcmin\, in Lupus 4. On the other hand, at 3 mm the telescope beam is significantly smaller (35\arcsec) and we mapped only the regions were NH$_3$ (1,1) emission was detected. Note that the Lupus 4 cloud was not mapped at 3 mm because we did not have enough time. The scanned regions were covered by mini maps of 5$\arcmin \times 5 \arcmin$ and 18$\arcmin \times 18 \arcmin$ for the observations at 3 and 12 mm, respectively. Each mini map was scanned twice in orthogonal directions in order to minimize artificial stripes and reduce noise level.

Data reduction was performed using the ATNF dedicated packages Livedata and Gridzilla\footnote
{http://www.atnf.csiro.au/computing/software/}. Livedata performs a bandpass calibration and baseline fitting while Gridzilla regrids and combines the data from multiple scanning directions and mini maps onto a single data cube. The data was Gaussian smoothed so that the effective spatial resolution of the final maps is 46\arcsec\, at 3 mm and 2\arcmin\, at 12 mm. 

\begin{figure*}
\includegraphics[width=8cm,angle=-90]{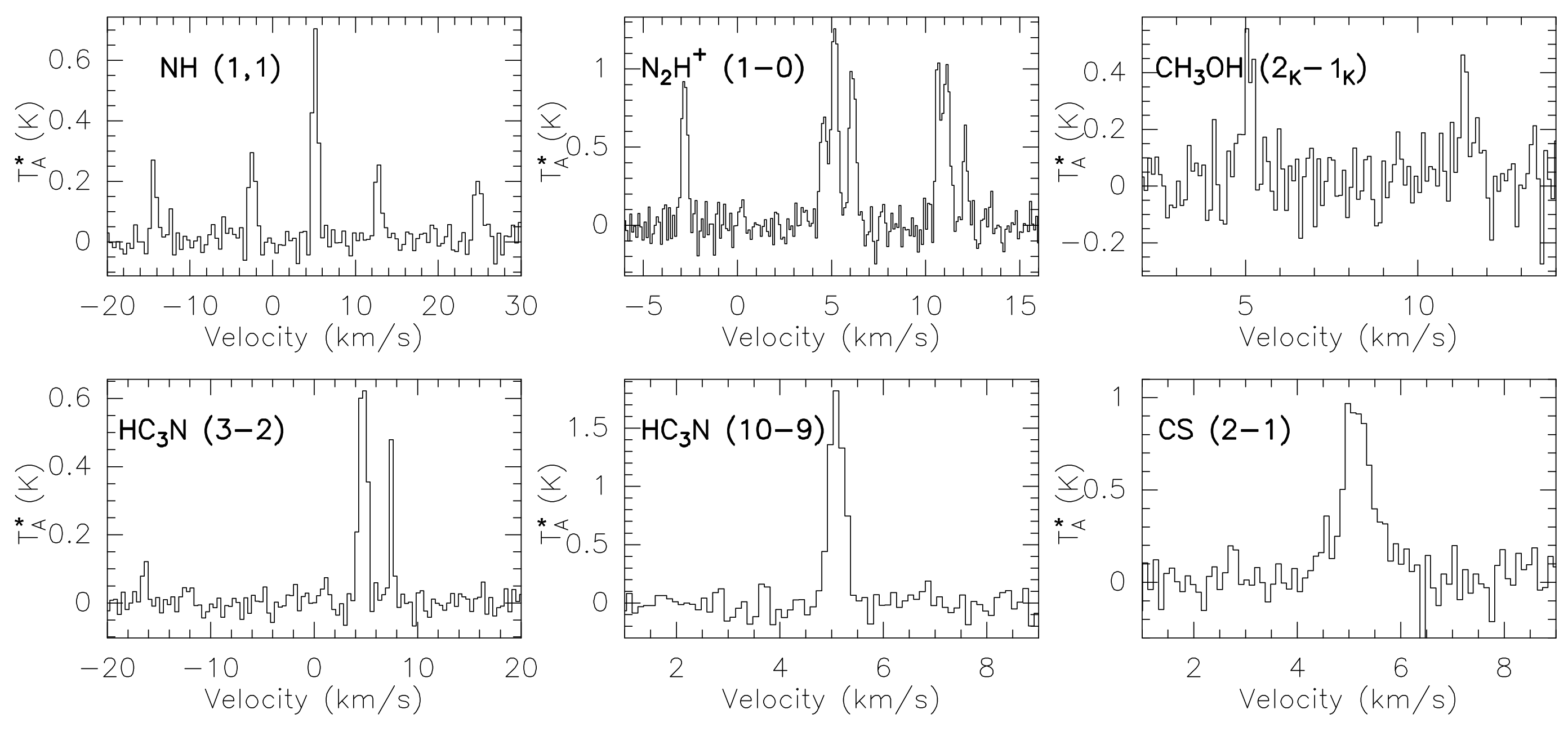}
\caption{Spectra of the observed transitions towards the source Lup1 C4.}
\label{spettri}
\end{figure*}

At 12 mm, we detected the NH$_3$ (1,1) transition at 23.694 GHz with the 4 satellites and the main component of the higher excitation transition NH$_3$ (2,2) at 23.722 GHz (only towards the sources with bright (1,1) emission), no NH$_3$ (3,3) emission was detected. At 12 mm we also detected two hyperfine transitions of the HC$_3$N (3-2) line, the main component  F=4$\rightarrow$3 at 27.294314 GHz blended with the second component F=3$\rightarrow$2  and the third component F=2$\rightarrow$1 at 27.294065 GHz, a fourth component F=2$\rightarrow$2 at 27.296230 GHz is detected only towards the brightest core in Lupus 1.
At 3 mm we detected the CS (2-1) line at 97.981 GHz, two methanol transitions the CH$_3$OH (2$_0$-1$_0$)A$^+$ at 96.741 GHz and CH$_3$OH (2$_{-1}$-1$_{-1}$)E at 96.739 GHz, the HC$_3$N (10-9) at 90.979 GHz and all the 7 hyperfine components of the N$_2$H$^+$ (1-0) transition around 93.174 GHz. An example of the spectra of the observed lines in Lupus 1 is given in Fig. \ref{spettri}.

A total of 15 high density gas cores have been detected in the three clouds. In Table \ref{coord} we give the coordinates and the size of the cores; these data have been derived in the 3 mm maps by using the tracer where the clump is best defined. The coordinates of the cores derived in the 3 mm maps are consistent with the peaks of the emission in the 12 mm maps, considering the larger beam of the 12 mm maps.

\begin{table*}
 \centering
 \begin{minipage}{140mm}
  \caption{List of the dense cores, their coordinates, sizes and kinetic temperatures.}
  \begin{tabular}{@{}rlcccc@{}}
  \hline
   Index  & Core identifier &  R.A. (J2000)  & Dec (J2000) & size (\arcsec) & T$_{kin}$ (K)\\
   1      & Lup1 C1    & 15 42 23.0 & -34 09 40.0 & 90  & ...\\
   2      & Lup1 C2    & 15 42 46.0 & -33 53 30.0 & 82  & ... \\
   3      & Lup1 C3    & 15 42 47.5 & -34 08 00.0 & 97  & ... \\
   4      & Lup1 C4    & 15 43 00.0 & -34 09 09.0 & 73  & 12.6\\
   5      & Lup1 C5    & 15 44 43.0 & -34 20 25.0 & 149 & ...\\
   6      & Lup1 C6    & 15 45 00.0 & -34 17 15.0 & 95  & 12.3\\
   7      & Lup1 C7    & 15 45 17.0 & -34 17 00.0 & 79  & $<$12.1\\
   8      & Lup1 C8    & 15 45 30.0 & -34 24 00.0 & 104 & ...\\
 \hline
   9      & Lup3 C1    & 16 08 49.7 & -39 07 20.0 & 43  & ...\\
  10      & Lup3 C2    & 16.09 16.5 & -39 07 13.0 & 57  & ...\\
  11      & Lup3 C3    & 16 09 19.0 & -39 04 44.0 & 94  & 12.6\\
  12      & Lup3 C4    & 16 09 23.0 & -39 06 54.0 & 36  & ...\\
  13      & Lup3 C5    & 16 09 38.5 & -39 05 00.0 & 62  & ... \\
 \hline
  14      & Lup4 C1    & 16 00 55.0 & -42 03 00.0 & 112 & 12.6\\
  15      & Lup4 C2    & 16 01 32.0 & -41 52 00.0 & 118 & 12.3\\
\hline
\end{tabular}
\end{minipage}
\label{coord}
\end{table*}

\subsection{Column densities}

In order to estimate the column densities of the observed species towards the cores, we defined polygons around each core at a level of 50 \% of the peak antenna temperature of the species where the core is best defined and extracted the averaged spectrum inside the defined polygon for all the species. However, because of the large difference between the HPBW of the 3 and 12 mm observations, we used different polygons for the two sets of lines.
The extracted spectra have been fitted with the Gaussian or the HFS (Hyperfine Structure) method of the GILDAS\footnote{http://www.iram.fr/IRAMFR/GILDAS/} software, developed at the IRAM and Observatore de Grenoble. The line parameters derived by the Gaussian fitting are listed in Table \ref{gauss} for the CS (2-1), CH$_3$OH (2$_0$-1$_0$)A$^+$ and CH$_3$OH (2$_{-1}$-1$_{-1}$)E and in Table \ref{gauss2} for NH$_3$ (2,2) and HC$_3$N (10-9) lines. 
The \ndueh (1-0), NH$_3$ (1,1) and \hctren (3-2) transitions show the hyperfine structure and these lines have been fitted using the method HFS of the GILDAS software. This method assumes that all the hyperfine components have the same excitation temperature and width, and that their separation is fixed to the laboratory value. The fitting provides an estimate of the total optical depth of the lines and the excitation temperature. The fitting results are reported in Table \ref{gauss2} and \ref{hyperfine}.

\begin{table*}
\centering
\caption{Line parameters derived from Gaussian fitting the CS (2-1), CH$_3$OH (2$_0$-1$_0$)A$^+$ and CH$_3$OH (2$_{-1}$-1$_{-1}$)E lines. Parameters without errors were fixed in the line fitting.}
\begin{tabular}{@{}cccccccccc@{}}
Core    &  \multicolumn{3}{c}{CS (2-1)} & \multicolumn{3}{c}{CH$_3$OH (2$_0$-1$_0$)A$^+$} & \multicolumn{3}{c}{CH$_3$OH (2$_{-1}$-1$_{-1}$)E} \\
& $v$ & $\Delta v$ & $\int{T_{MB} dv}$ & $v$ & $\Delta v$ & $\int{T_{MB} dv}$ & $v$ & $\Delta v$ & $\int{T_{MB} dv}$ \\
& \kms & \kms & K \kms& \kms & \kms & K \kms & \kms & \kms & K \kms\\
\hline
Lup1 C1 & 5.54$\pm$0.04 & 0.6$\pm$0.1   & 1.36$\pm$0.07 & 5.54$\pm$0.04 & 0.6$\pm$0.1   & 0.52$\pm$0.07 & 11.75$\pm$0.06 & 0.5$\pm$0.1   & 0.29$\pm$0.06 \\
Lup1 C2 & 4.63$\pm$0.03 & 1.15$\pm$0.08 & 1.52$\pm$0.09 & 4.72$\pm$0.04 & 0.5$\pm$0.1   & 0.45$\pm$0.07 & 10.89$\pm$0.04 & 0.37$\pm$0.08 & 0.27$\pm$0.06 \\
Lup1 C3 & 5.14$\pm$0.03 & 0.99$\pm$0.08 & 1.65$\pm$0.1  & 5.03$\pm$0.03 & 0.52$\pm$0.08 & 0.62$\pm$0.08 & 11.31$\pm$0.03 & 0.35$\pm$0.07 & 0.43$\pm$0.07 \\
Lup1 C4 & 5.16$\pm$0.02 & 0.72$\pm$0.07 & 1.57$\pm$0.1  & 5.09$\pm$0.03 & 0.4$\pm$0.1   & 0.46$\pm$0.07 & 11.38$\pm$0.06 & 0.6$\pm$0.2   & 0.48$\pm$0.08 \\
Lup1 C6	& 4.94$\pm$0.02 & 0.54$\pm$0.06 & 1.10$\pm$0.07 & 4.90$\pm$0.02 & 0.43$\pm$0.06 & 0.53$\pm$0.05 & 11.10$\pm$0.03 & 0.42$\pm$0.07 & 0.43$\pm$0.05 \\
 ''     & 4.2$\pm$0.2   & 1.1$\pm$0.4   & 0.64$\pm$0.07 & ... & ... & ...  & ...  & ... & ... \\
Lup1 C7	& 5.01$\pm$0.02 & 0.87$\pm$0.06 & 1.79$\pm$0.09 & 5.06$\pm$0.04 & 0.7$\pm$0.1   & 0.50$\pm$0.06 & 11.27$\pm$0.06 & 0.9$\pm$0.1   & 0.34$\pm$0.06 \\
Lup1 C8	& 4.36$\pm$0.03 & 0.62$\pm$0.07 & 0.84$\pm$0.07 & 4.29$\pm$0.04 & 0.34$\pm$0.09 & 0.22$\pm$0.05 & 10.59$\pm$0.02 & 0.17$\pm$0.04 & 0.16$\pm$0.04 \\

Lup3 C1 & 4.16$\pm$0.02 & 0.78$\pm$0.05 & 1.87$\pm$0.1  & 4.01$\pm$0.07 & 0.6$\pm$0.2  & 0.48$\pm$0.09 & 10.44$\pm$0.07 & 0.5 $\pm$0.1 & 0.33$\pm$0.09 \\
Lup3 C2 & 4.71$\pm$0.04 & 0.48$\pm$0.07 & 1.07$\pm$0.08 & 4.6$\pm$0.1	& 0.4$\pm$0.2  & 0.35$\pm$0.07 & 10.80$\pm$0.05 & 0.4	      & 0.28$\pm$0.07 \\
 ''     & 4.15$\pm$0.05 & 0.63$\pm$0.09 & 1.49$\pm$0.08 & 4.1$\pm$0.1	& 0.4$\pm$0.2  & 0.44$\pm$0.07 & 10.30$\pm$0.07 & 0.4	      & 0.39$\pm$0.07 \\
Lup3 C3	& 4.65$\pm$0.01 & 0.44$\pm$0.03 & 0.82$\pm$0.06 & ... & ... & ...  & ...  & ... & ...  \\
Lup3 C4 & 4.60$\pm$0.03 & 0.49$\pm$0.06 & 1.43$\pm$0.1  & 4.55$\pm$0.04 & 0.5          & 0.43$\pm$0.08 & 10.71$\pm$0.06 & 0.5	      & 0.32$\pm$0.08 \\
  ''    & 4.07$\pm$0.04 & 0.45$\pm$0.06 & 1.17$\pm$0.1  & 4.02$\pm$0.05 & 0.45         & 0.64$\pm$0.08 & 10.26$\pm$0.07 & 0.45	      & 0.50$\pm$0.08 \\
Lup3 C5	& 4.73$\pm$0.01 & 0.53$\pm$0.03 & 1.05$\pm$0.06 & 4.76$\pm$0.03 & 0.44$\pm$0.08& 0.36$\pm$0.05 & 11.00$\pm$0.04 & 0.49$\pm$0.08 & 0.37$\pm$0.05 \\
\hline
\end{tabular}
\label{gauss}
\end{table*}

\begin{table*}
\centering
  \caption{Line parameters derived from Gaussian fitting the HC$_3$N (10-9) and NH$_3$ (2,2) lines and from  HFS fitting the NH$_3$ (1,1) line. Parameters without errors were fixed in the line fitting.}
\begin{tabular}{@{}ccccccccccc@{}}
Core    &  \multicolumn{3}{c}{HC$_3$N (10-9)}&  \multicolumn{4}{c}{NH$_3$ (1,1)} & \multicolumn{3}{c}{NH$_3$ (2,2)}\\
        & $v$ & $\Delta v$ & $\int{T_{MB} dv}$ & $v$ & $\Delta v$ & $\tau_{main}$ & T$_{ex}$& $v$ & $\Delta v$ & $\int{T_{MB} dv}$  \\
        & \kms & \kms & K \kms & \kms &\kms & & K & \kms & \kms & K \kms \\
\hline
Lup1 C3 & 5.08$\pm$0.01 & 0.32$\pm$0.02 & 1.19$\pm$0.06&  ...  	 & ...  	   & ...	 & ...        & ... & ... & ...\\
Lup1 C4 & 5.10$\pm$0.01 & 0.34$\pm$0.02 & 1.51$\pm$0.06& 5.16$\pm$0.02 & 0.74$\pm$0.01 & 1.3$\pm$0.3 & 3.9$\pm$1.2& 5.29$\pm$0.1 & 0.8$\pm$0.2 & 0.13$\pm$0.03 \\
Lup1 C6	& 4.91$\pm$0.01 & 0.42$\pm$0.02 & 1.45$\pm$0.05& 4.92$\pm$0.03 & 0.71$\pm$0.02 & 2.5$\pm$0.5 & 3.5$\pm$1.0& 4.75$\pm$0.1 & 0.7$\pm$0.3 & 0.11$\pm$0.04 \\
Lup1 C7	& 5.04$\pm$0.03 & 0.45$\pm$0.06 & 0.36$\pm$0.05& 5.01$\pm$0.04 & 0.71$\pm$0.05 & 0.2$\pm$0.1 & 6.5$\pm$4.7& ... & ... & $<$0.06\\
Lup1 C8	& 4.37$\pm$0.02 & 0.41$\pm$0.04 & 0.57$\pm$0.05& ...  	 & ...  	   & ...	 & ...        & ... & ... & ...\\
Lup3 C2 & 4.75$\pm$0.05 & 0.4$\pm$0.1   & 0.33$\pm$0.07& ...  	 & ...             & ...         & ...        & ... & ... & ...\\
Lup3 C3	& 4.69$\pm$0.02 & 0.30$\pm$0.04 & 0.35$\pm$0.05& 4.71$\pm$0.02 & 0.71$\pm$0.01 & 1.1$\pm$0.3 & 4.2$\pm$1.4& 4.9$\pm$0.1 & 0.7 & 0.09$\pm$0.03\\
Lup3 C5	& 4.85$\pm$0.01 & 0.39$\pm$0.03 & 0.80$\pm$0.07& ...  	 & ...  	   & ...	 & ...        & ... & ... & ...\\
Lup4 C1 & ...           & ...           & ...          & 4.11$\pm$0.03 & 0.71$\pm$0.04 & 0.9$\pm$0.5 & 3.6$\pm$2.7& 3.7$\pm$0.2 & 0.8$\pm$0.5 & 0.07$\pm$0.03\\
Lup4 C2 & ...           & ...           & ...          & 4.06$\pm$0.04 & 0.71$\pm$0.05 & 0.1$\pm$0.06& 7.6$\pm$5.3& 4.6$\pm$0.1 & 0.8$\pm$0.3 & 0.10$\pm$0.03\\
\hline
\end{tabular}
\label{gauss2}
\end{table*}

\begin{table*}
 \centering
  \caption{Line parameters derived from HFS fitting the \ndueh(1-0) and \hctren (3-2) lines.}
\begin{tabular}{@{}lcccccccc@{}}
Core    &  \multicolumn{4}{c}{\ndueh(1-0)} &   \multicolumn{4}{c}{\hctren (3-2) }\\
& $v$  & $\Delta v$ & $\tau_{tot}$ & T$_{ex}$  & $v$  & $\Delta v$ & $\tau_{tot}$ & T$_{ex}$ \\
& \kms & \kms & & K  & \kms & \kms & & K \\
\hline
Lup1 C3 & 5.14$\pm$0.01   & 0.29$\pm$0.02 & 3.2$\pm$0.6  & 5$\pm$1 & ... & ... & ... & ... \\
Lup1 C4 & 5.169$\pm$0.005 &0.309$\pm$0.007& 12.0$\pm$0.5 & 5$\pm$2 & 4.88$\pm$0.01 & 0.62$\pm$0.02 & 3.6$\pm$0.4& 3.4$\pm$0.6 \\
Lup1 C5 & ...             & ...           & ...          & ...     & 4.77$\pm$0.03 & 0.71$\pm$0.08 & 0.8$\pm$0.6& 4$\pm$3 \\
Lup1 C6	& 4.981$\pm$0.009 & 0.39$\pm$0.02 & 11$\pm$3     & 4$\pm$1 & 4.63$\pm$0.02 & 0.71$\pm$0.05 & 2.3$\pm$0.6& 4$\pm$1 \\
Lup1 C7	& 5.007$\pm$0.008 & 0.36$\pm$0.02 & 8$\pm$2	 & 4$\pm$2 & ... & ... & ... & ... \\
Lup1 C8	& ...             & ...           & ...          & ...     & 4.20$\pm$0.02 & 0.62$\pm$0.09 & 5.7$\pm$0.2& 3.1$\pm$0.3 \\
Lup3 C1 & 4.08$\pm$0.04   & 0.51$\pm$0.09 & 2$\pm$1      & 4$\pm$3 & ... & ...& ... & ... \\
Lup3 C2 & 4.694$\pm$0.001 & 0.27$\pm$0.03 & 9$\pm$1      & 3.3$\pm$0.7& ... & ...& ... & ... \\
Lup3 C3	& 4.731$\pm$0.007 & 0.34$\pm$0.02 & 2$\pm$1      & 7$\pm$4 & 4.32$\pm$0.08 & 0.6$\pm$0.1   & 1.1$\pm$0.8 & 3$\pm$2 \\
Lup3 C4 & ...             & ...           & ...          & ...          & 4.31$\pm$0.07 & 0.65     & ...	 & ...        \\
Lup3 C5	& 4.84$\pm$0.01   & 0.35$\pm$0.03 & 3.5$\pm$0.9  & 5$\pm$1 & 4.61$\pm$0.04 & 0.65$\pm$0.09 & 2$\pm$1 & 4$\pm$3 \\
Lup4 C1 & ...             & ...           & ...          & ...     & 3.83$\pm$0.02 & 0.62$\pm$0.05 & 0.8$\pm$0.5 & 4$\pm$3 \\
Lup4 C2 & ...             & ...           & ...          & ...     & 3.87$\pm$0.03 & 0.73$\pm$0.09 & 3$\pm$1 & 3$\pm$2 \\
\hline
\end{tabular}
\label{hyperfine}
\end{table*}

We used different methods for the calculation of the column densities of the observed species. 
For the \ndueh (1-0), NH$_3$ (1,1) and \hctren (3-2) transitions we derived a quite reliable estimate of the column density since we have a direct estimate of the line opacity and the excitation temperature from fitting the hyperfine structure under the assumption that the source fills the beam. On the other hand, for \hctren (10-9), CS (2-1) and the two CH$_3$OH lines we do not know the line optical depth or the temperature. Therefore we calculated the column density under the optical thin and LTE approximation, assuming the rotational temperature derived from NH$_3$ or, for the cores where there is not this estimate, assuming the typical value derived in dense cores, T=10 K \citep{tafalla02}. These values should be considered lower limits. The column densities are reported in Table \ref{column} and the details of the calculation are given in Appendix A. The error associated to the column density is derived from the propagation of the errors of the parameters derived by the fitting of the lines. The three species showing the hyperfine components and fitted with the HFS method, have  higher errors mainly due to the uncertainties associated to the line opacity an excitation temperature and in those cores where the fainter hyperfine components were not detected with high signal to noise, the uncertainty is as high as 70\%. On the other hand, the column densities derived from the lines fitted with a single Guassian have a lower relative error but they have also an uncertainty, not quantifiable, related to the assumptions made on their optical depth and excitation temperature.

For \hctren\,we derived the total column density from the two observed lines independently (column 6 and 7 in Table \ref{column}). The estimates derived from the \hctren (10-9) line are lower of a factor $\sim$ 4 (except for Lup3 C3 where the factor is 14) with respect to the estimates derived by the (3-2) line. 
This systematic trend suggests that probably the (10-9) is not optically thin as we assumed or there could be an effect of beam dilution. It is worth nothing that the two lines have been observed with different beams and so the column densities are averaged over different areas.

Also for \chtreoh we derived the total column density from the two observed lines independently (column 3 and 4 in Table \ref{column}), and we found that the total methanol column density derived from the (2$_{-1}$-1$_{-1}$) E line is systematically higher by a small factor (between 1.3 and 2.5) compared to the value derived from the (2$_0$-1$_0$) A$^+$ line, even if in a few cores the two measures agree within the errors. Also in this case this little discrepancy could be due to opacity effects.

\begin{table*}
  \caption{Total column densities of the observed species.}
\begin{tabular}{@{}lccccccc@{}}
\hline
Core & N(CS) & N(\chtreoh)$^a$ & N(\chtreoh)$^b$ & N(\ndueh) & N(\hctren)$^c$ & N(\hctren)$^d$ & N(NH$_3$)$^f$ \\
     & 10$^{12}$ cm$^{-2}$ & 10$^{13}$ cm$^{-2}$ & 10$^{13}$ cm$^{-2}$ & 10$^{12}$ cm$^{-2}$ & 10$^{13}$ cm$^{-2}$ & 10$^{13}$ cm$^{-2}$ & 10$^{13}$ cm$^{-2}$ \\
Lup1 C1 & 5.2$\pm$0.3 & 2.5$\pm$0.3 & 3.3$\pm$0.7 & ... & ... & ... & ...\\
Lup1 C2 & 5.8$\pm$0.4 & 2.2$\pm$0.3 & 3.0$\pm$0.7 & ... & ... & ... & ... \\
Lup1 C3 & 6.3$\pm$0.5 & 3.0$\pm$0.4 & 4.8$\pm$0.8 & 2.5$\pm$0.7 & 2.4$\pm$0.5 & ... & ... \\
Lup1 C4 & 6.3$\pm$0.5 & 3.0$\pm$0.5 & 6.5$\pm$1.1 &11$\pm$1 & 2.5$\pm$0.5 & 11$\pm$1    & 11$\pm$4 \\
Lup1 C5 & ... & ... & ... & ... & ...                                     & 3$\pm$2 &  ...\\
Lup1 C6 & 4.4$\pm$0.4 & 3.4$\pm$0.3 & 5.7$\pm$0.7 & 8$\pm$3 & 2.4$\pm$0.5 & 9$\pm$3 & 17$\pm$6 \\
Lup1 C7 & 6.9$\pm$0.5 & 2.4$\pm$0.3 & 3.8$\pm$0.7 & 6$\pm$2 & 0.7$\pm$0.2 & ...     & 2.6$\pm$0.2 \\
Lup1 C8 & 3.2$\pm$0.3 & 1.1$\pm$0.2 & 1.8$\pm$0.5 & ...     & 1.1$\pm$0.3 & 16$\pm$2    & ... \\
Lup3 C1 & 7.2$\pm$0.4 & 2.3$\pm$0.4 & 3.7$\pm$1.0 & 2$\pm$1 & ...& ... & ... \\
Lup3 C2 & 4.1$\pm$0.4 & 1.7$\pm$0.3 & 3.1$\pm$0.8 & 3.3$\pm$0.8 & 0.7$\pm$0.2 & ... & ... \\
Lup3 C3 & 3.3$\pm$0.3 & ...     & ...             & 3$\pm$2 & 0.6$\pm$0.1 & 3$\pm$2 & 9$\pm$4 \\
Lup3 C4 & 5.5$\pm$0.4 & 2.1$\pm$0.4 & 3.6$\pm$0.9 & ... & ... & ...     & ... \\
Lup3 C5 & 4.0$\pm$0.3 & 1.7$\pm$0.2 & 4.2$\pm$0.6 & 2.7$\pm$0.8 & 1.6$\pm$0.3 & 8$\pm$4 & ... \\
Lup4 C1 & ... & ... & ... & ... & ...                                         & 3$\pm$2 & 6$\pm$5 \\
Lup4 C2 & ... & ... & ... & ... & ...                                         & 9$\pm$3 & 3$\pm$2 \\
\hline
\end{tabular}
\\
$^a$ derived from the (2$_0$-1$_0$) A$^+$ line \\
$^b$ derived from the (2$_{-1}$-1$_{-1}$)E line \\
$^c$ derived from the (10-9) line\\
$^d$ derived from the (3-2) line\\
$^f$ derived from the (1,1) line \\
\label{column}
\end{table*}

\subsection{Temperature}

In five out of the six cores where the NH$_3$ (1,1) line is detected, we detect also the (2,2) transition so that we can estimate the kinetic temperature (for the sixth core we derived the upper limit). The values are reported in Table \ref{coord} and the details of the calculation are given in Appendix B. For all the cores we derived kinetic temperatures between 12 and 13 K. These values are usually measured at the edge of the cold prestellar cores, indicating the relative youthfulness of the observed cores. However, high spatial resolution observations have shown that the kinetic temperature of dense cores typically drops at or below 10 K at the core centre (e.g. \citealt{tafalla04}; \citealt{crapsi07}). On the other hand, a kinetic temperature around 12 K is measured in dense cores associated with protostars \citep{foster09}.
Therefore, we mostly trace the outskirts of dense cores or, alternatively the Lupus cores where we detect ammonia are more evolved than other well known dense starless cores, being associated with a protostar.

\section{Results}

The line intensity maps of the observed species show that the CS (2-1) line traces the large scale structure of the dense gas inside the clouds and it can be also used to probe the kinematics of the gas (\citealt{testi00}; \citealt{olmi02}). On the other hand, NH$_3$, HC$_3$N and N$_2$H$^+$ trace the high density condensations. 
In general towards the dense cores we found that the line velocities and lines widths derived from the different species agree within one element of spectral resolution, indicating that they are tracing the same gas.

In the following subsections we report the results of each cloud separately.

\subsection{Lupus 1}

\subsubsection{Large scale structure}

\begin{figure*}
\includegraphics[width=15cm,angle=-90]{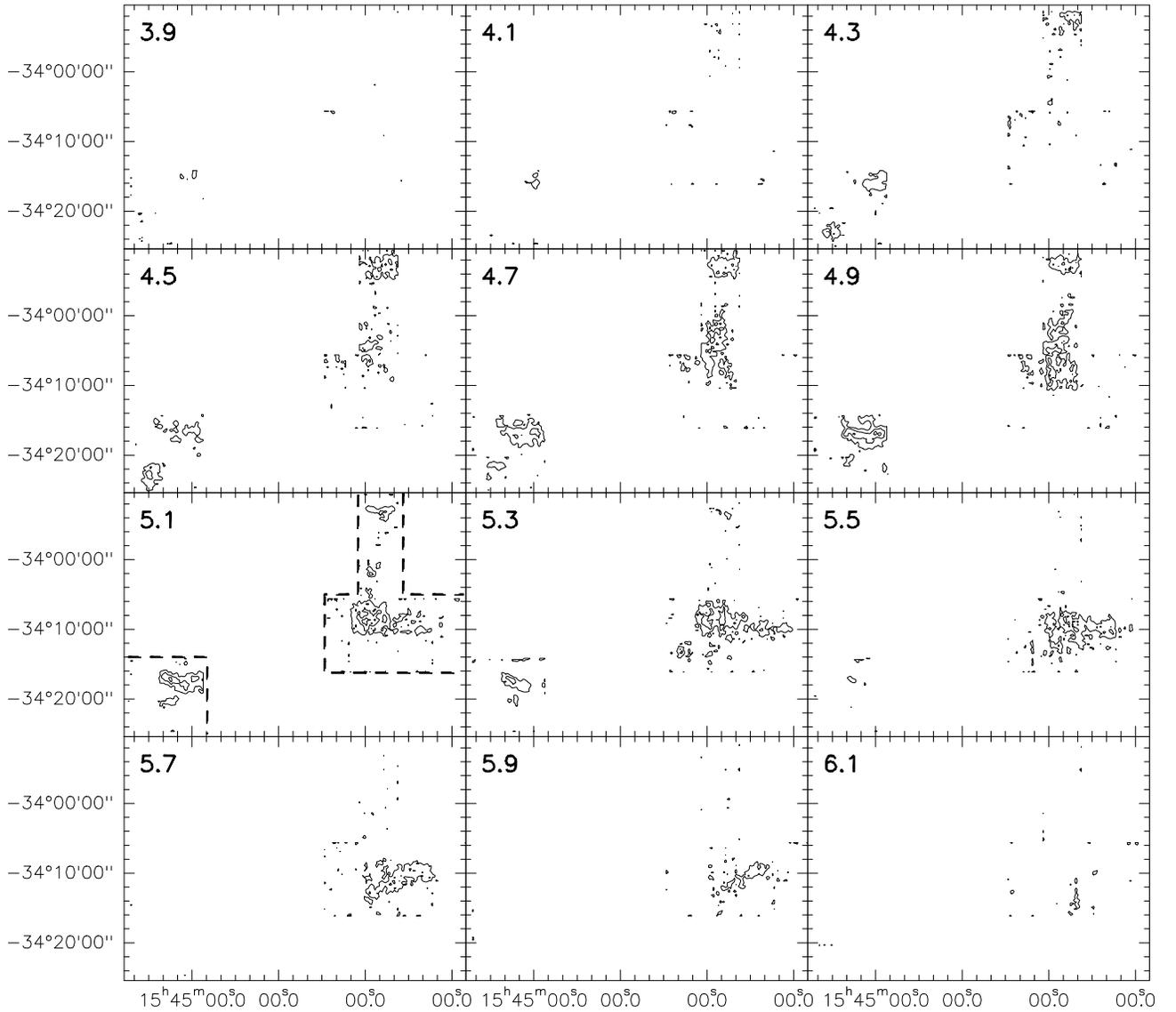}
\caption{CS (2-1) channel map of Lupus 1; first contour and contour steps are 0.1 K \kms. The dashed boxes in the 5.1 \kms\, panel indicate the mapped zone.}
\label{lup1_cs}
\end{figure*}

Examining the channel map for the CS (2-1) line (Fig. \ref{lup1_cs}) one can see the presence of velocity gradients in the molecular cloud with increasing gas velocity from north to south and from east to west. The gas in the northern part is at lower velocity starting from $\sim$ 4.3 \kms, the velocity increases till $\sim$5.9 \kms\, towards the south. Along the southern part of the cloud we observe a horizontal velocity gradient, with velocity increasing from $\sim$  3.9 \kms\, at east to 6.0 \kms\, at west.  Also the peak of the CS line, i.e. the v$_{lsr}$, ranges from 4.6 \kms\, to 5.5 \kms\, moving from north to south and from 4.4 to 5.5 moving from south-east to south-west.
Moreover towards the source Lup1 C6, the CS (2-1) line shows a double peak, one at 4.9 \kms\, and one at 4.2 \kms, that indicates that two velocity components are present along this line of sight. However, we cannot rule out the possibility that this is a self absorption of the CS (2-1) line since we do not detect the C$^{34}$S (2-1) line.

\subsubsection{Dense cores}

\begin{figure*}
\includegraphics[width=9cm,angle=-90]{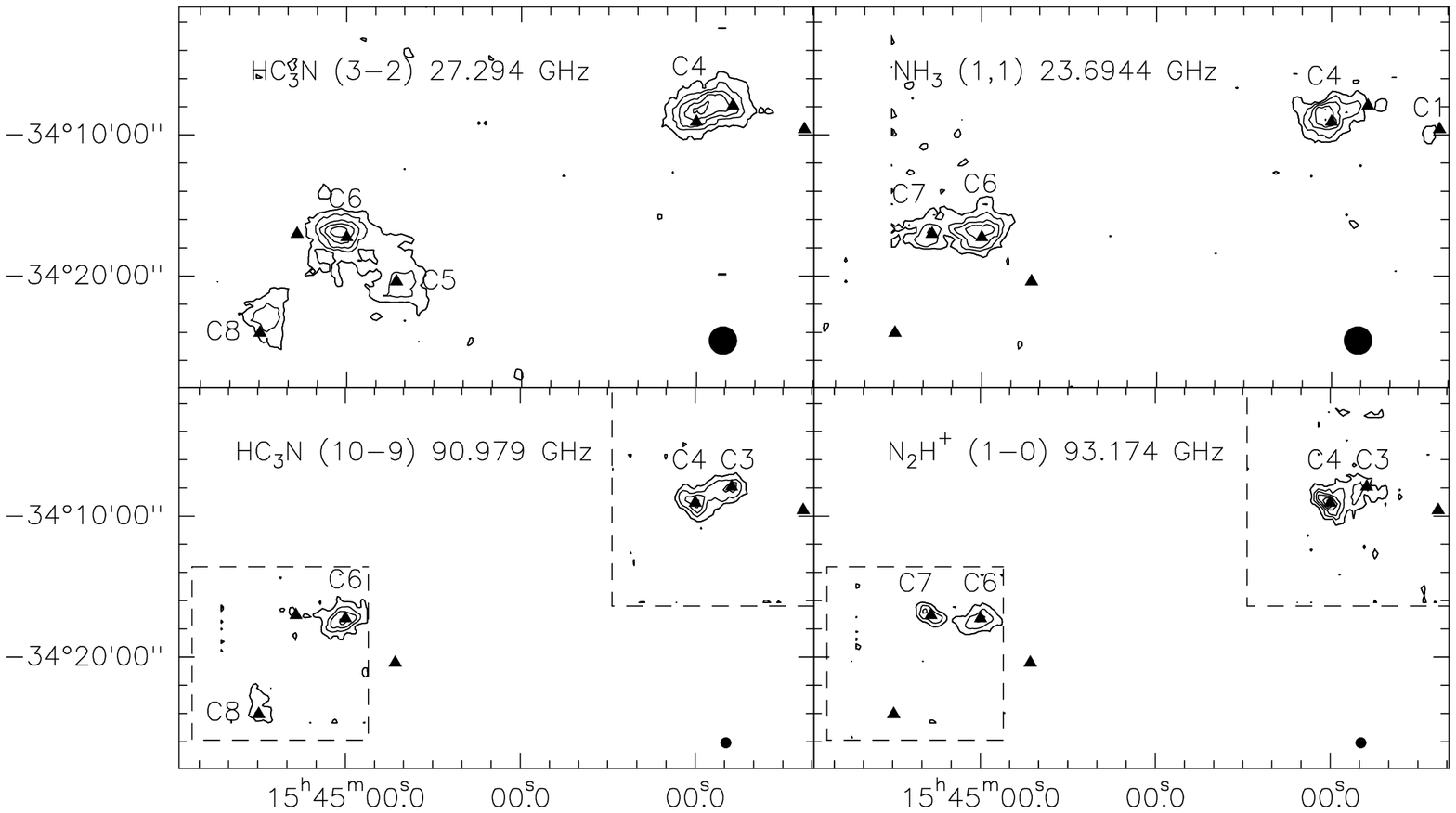}
\caption{Line maps of Lupus 1 in NH$_3$ (1,1), N$_2$H$^+$ (1-0), HC$_3$N (3-2) and \hctren (10-9). The dashed boxes in the lower panels indicate the mapped zone in N$_2$H$^+$ (1-0), HC$_3$N (10-9) (in fact the cloud has been partially mapped in these two lines). Contours are as follow: for \hctren (3-2) first contour 0.2 K \kms, step 0.2 K \kms; for NH$_3$ (1,1) first contour 0.3 K \kms, step 0.15 K \kms; for \hctren (10-9) first contour 0.2 K \kms, step 0.2 K \kms; for N$_2$H$^+$ (1-0) first contour 0.3 K \kms, step 0.3 K \kms. The filled circles represent the HPBW of the maps and the triangles mark the position of the dense cores.}
\label{lup1}
\end{figure*}

\begin{figure}
\includegraphics[width=4cm,angle=-90]{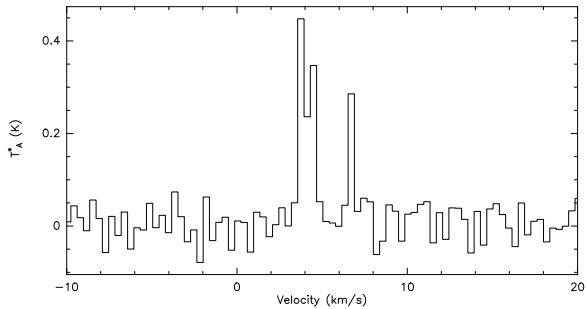}
\caption{Spectrum of the HC$_3$N (3-2) line toward the core Lup1 C8.}
\label{hc3n_c8}
\end{figure}

In Fig. \ref{lup1} we show the maps of the Lupus 1 central region in NH$_3$ (1,1), N$_2$H$^+$ (1-0), HC$_3$N (3-2) and \hctren (10-9). In these tracers we detect overall 8 cores. The coordinates and the sizes of the cores are given in Table \ref{coord} and the column densities of the observed species towards the cores are given in Table \ref{column}. The cores are better defined in the higher spatial resolution maps, i.e. \ndueh (1-0) and \hctren (10-9). Cores Lup1 C1 and Lup1 C2 are visible in the CS and \chtreoh maps, only marginally in NH$_3$, but not in the other species. Core Lup1 C7 is not detected in \hctren (3-2) and only marginally in the higher (10-9) line. On the contrary, Lup1 C8 is detected in \hctren\,and CS but not in NH$_3$ and \ndueh. Similarly, Lup1 C5 is detected in \hctren (3-2) but not in NH$_3$; it has not been mapped in \ndueh.
The  HC$_3$N (3-2) main component in Lup1 C8 show an asymmetric line profile (see Fig. \ref{hc3n_c8}) that could be a sign of infall. However, at the spectral resolution of the measure, the line profile is spread over only 3 spectral pixels, making difficult the infall interpretation. In fact, it could be a self--absorption of the line since toward that core we derived the higher optical depth and column density of HC$_3$N.

\subsection{Lupus 3}

\subsubsection{Large scale structure}
\begin{figure*}
\includegraphics[width=8cm,angle=-90]{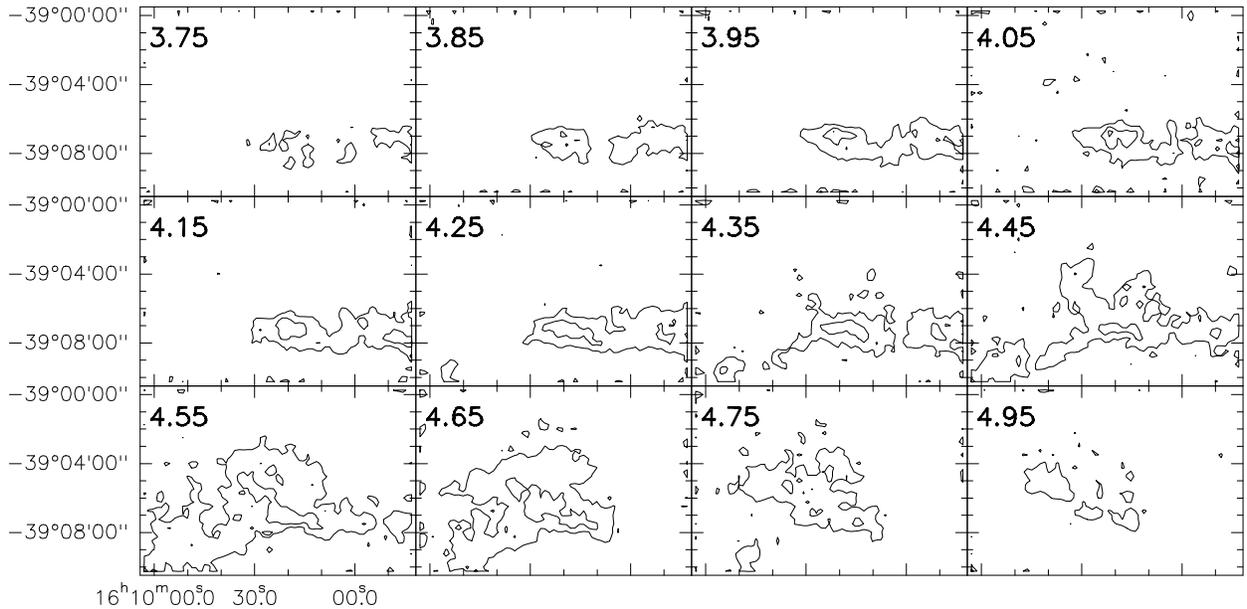}
\caption{CS (2-1) channel map of Lupus 3; first contour and contour steps are 0.1 K \kms.}
\label{lup3_cs}
\end{figure*}

Looking at the channel map of the CS (2-1) line (Fig. \ref{lup3_cs}) one can see the presence of different velocity components: an horizontal filament in the south with velocity between 3.7 and 4.3 \kms, a central component with velocity around  4.6 \kms and an oblique filament at east with velocity up to 5 \kms. In particular in the southern part of the map there are clearly two velocity components along the same line of sight one at v$_{lsr}\sim$ 4.2 \kms\, (the horizontal filament) and one at v$_{lsr}\sim$ 4.7 \kms\, (the central component) as indicated also from the double peak in the spectra of the CS (2-1) and \chtreoh (2$_K$-1$_K$) lines. It is worth nothing that in Lupus 3 the \chtreoh emission is present only in the southern horizontal filament and not in the central part of the cloud.

\subsubsection{Dense cores}

\begin{figure*}
\includegraphics[width=9cm,angle=-90]{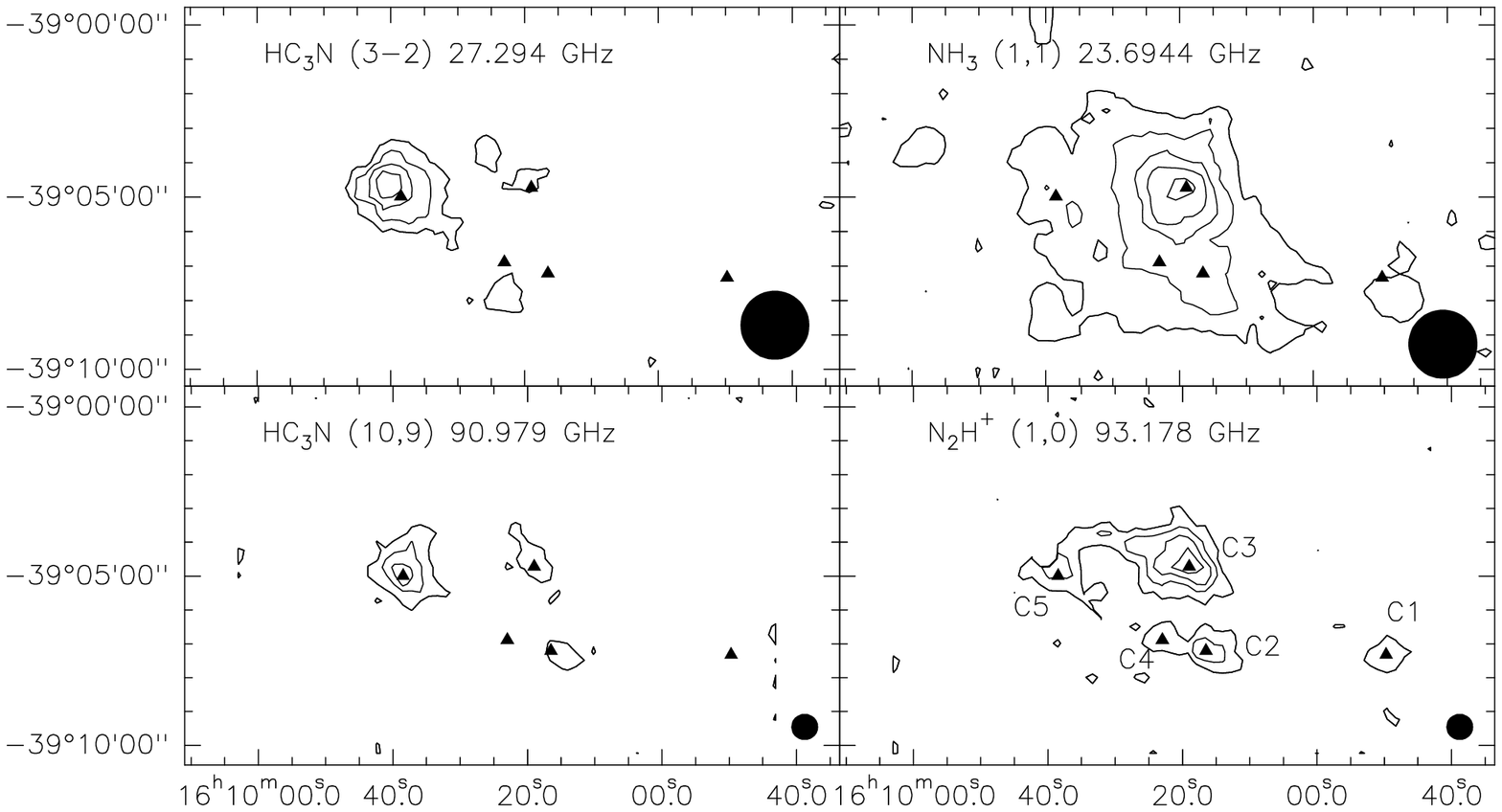}
\caption{Line maps of Lupus 3 in NH$_3$ (1,1), N$_2$H$^+$ (1-0), HC$_3$N (3-2) and \hctren (10-9). Contours are as follow: for \hctren (3-2) first contour 0.15 K \kms, step 0.15 K \kms; for NH$_3$ (1,1) first contour 0.18 K \kms, step 0.2 K \kms; for \hctren (10-9) first contour 0.15 K \kms, step 0.15 K \kms; for N$_2$H$^+$ (1-0) first contour 0.27 K \kms, step 0.2 K \kms. The filled circles represent the HPBW of the maps and the triangles mark the position of the dense cores.}
\label{lup3}
\end{figure*}

In Fig. \ref{lup3} we show the maps of the Lupus 3 central region in NH$_3$ (1,1), N$_2$H$^+$ (1-0), HC$_3$N (3-2) and \hctren (10-9). The cores are better defined in the higher spatial resolution maps, i.e. \ndueh (1-0) and \hctren (10-9), where we detected 5 dense cores. The coordinates and the sizes of the cores are given in Table \ref{coord} and the column densities of the observed species towards the cores are given in Table \ref{column}. In the two southern cores Lup3 C2 and Lup3 C4 two velocity components are detected in CS and \chtreoh; in Table \ref{column} we report only the column density of the more extended 4.7 \kms\, velocity component. The cores Lup3 C2, C3 and C5 are coincident with the three cores C, B and A respectively, detected in H$^{13}$CO$^+$ and the 1.2 mm continuum by \cite{tachihara07}, while Lup3 C1 is not detected in H$^{13}$CO$^+$ but is visible in their 1.2 mm continuum map. What \citet{tachihara07} call core C in the H$^{13}$CO$^+$ map is a quite elongated emission that in our \ndueh map is in fact composed by two different cores that we call Lup3 C2 and Lup3 C4. The two cores are also clearly visible in the methanol map. The methanol map correlates with the 1.2 mm map of \citet{tachihara07} in the southern horizontal filament.

The spectral fit of the CS emission show that in the Lup3 C2 and C4 positions two velocity components coexist along the same line of sight: one at 4.7 \kms\, and one at 4.2 \kms. It is worth nothing that  along the line of sight of the Lup3 C2 core we detect only the \hctren (10-9) line at 4.7 \kms, while along the line of sight of the Lup3 C4 core we detect only the lower energy transition J = 3 to 2 of the \hctren\, molecule at 4.3 \kms\, indicating that we are looking at two gas components with different v$_{lsr}$ and physical conditions along similar line of sight.

\subsection{Lupus 4}

\begin{figure}
\includegraphics[width=4.3cm,angle=-90]{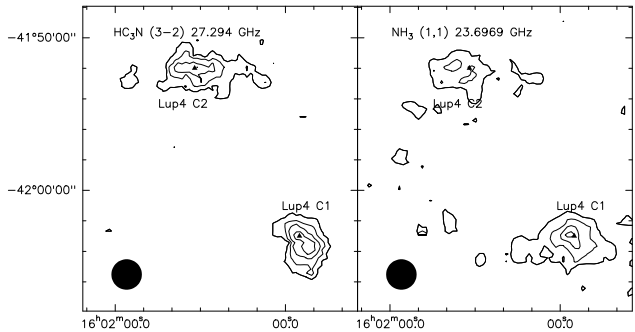}
\caption{Line maps of Lupus 4 in NH$_3$ (1,1) and HC$_3$N (3-2). Contours are as follow: for \hctren (3-2) first contour 0.18 K \kms, step 0.1 K \kms; for NH$_3$ (1,1) first contour 0.22 K \kms, step 0.15 K \kms. The filled circles represent the HPBW of the maps and the triangles mark the position of the dense cores.}
\label{lup4}
\end{figure}

In the Lupus 4 region only measurements at 12 mm were carried out. The NH$_3$ (1,1) and HC$_3$N (3-2) maps are shown in Fig. \ref{lup4}. Two dense cores have been identified: Lup4 C1 is bright in both lines, though it shows a different morphology in the two lines, while Lup4 C2 is fainter in ammonia. The coordinates and the sizes of the cores, as well as the kinetic temperature, are given in Table \ref{coord} and the column densities in Table \ref{column}.

\section{Chemical modeling}

From a first look at the line maps (Figs. \ref{lup1}, \ref{lup3} and \ref{lup4}) it is evident that the relative intensity of the lines of the three species \hctren, NH$_3$ and \ndueh changes significantly between the various cores: cores that are brighter in \hctren\,are fainter or undetected in NH$_3$ and \ndueh and vice versa.
Although previous studies suggested that the abundance ratio of NH$_3$ over \ndueh can be considered as a chemical clock since it is higher towards starless cores than towards protostellar cores (\citealt{caselli02a}; \citealt{aikawa05}; \citealt{busquet10}; \citealt{friesen10}), our maps show a strong anticorrelation between \hctren\, on one side and \ndueh and NH$_3$ on the other side. This behavior can be quantified by considering the ratio between the column densities that may reflect the ratio of the chemical abundances of the species if all lines are emitted by the same region. Therefore, we calculated the ratio HC$_3$N/\ndueh (Column 6/Column 5 of Table \ref{column}) and the ratio HC$_3$N/NH$_3$ (Column 7/Column 8 of Table \ref{column}), considering column densities derived from lines observed with similar angular resolution. We found indeed, that the two ratios change by about one order of magnitude between cores (see Table \ref{column_ratio}). In particular, the HC$_3$N/\ndueh ratio ranges from 1 to 10 and the HC$_3$N/NH$_3$ ratio ranges from 0.3 to 3. Even if the errors associated to the ratios are high, the observed trend is certainly significant for the HC$_3$N/\ndueh ratio while it is less so for the HC$_3$N/NH$_3$ ratio. However, we point out that there are 4 cores that emit in HC$_3$N, and not in NH$_3$, and 3 cores that emit in NH$_3$, and not in HC$_3$N clearly indicating a change in the chemical abundance ratio of the two species.

\begin{table}
  \caption{Column densities ratios.}
\begin{tabular}{@{}ccc@{}}
\hline
Core &  N(\hctren)$^a$/N(\ndueh) & N(\hctren)$^b$/N(NH$_3$)$^c$ \\
Lup1 C3 & 10$\pm$5 & ... \\
Lup1 C4 & 2$\pm$0.6 & 1$\pm$0.4 \\
Lup1 C6 & 3$\pm$2 & 0.5$\pm$0.3 \\
Lup1 C7 & 1.2$\pm$0.7 & ... \\
Lup3 C2 & 2$\pm$1 & ... \\
Lup3 C3 & 2$\pm$2  & 0.3$\pm$0.3 \\
Lup3 C5 & 6$\pm$3 & ... \\
Lup4 C1 & ... & 0.5$\pm$0.5 \\
Lup4 C2 & ... & 3$\pm$3 \\
\hline
\end{tabular}
\\
$^a$ derived from the (10-9) line\\
$^b$ derived from the (3-2) line\\
$^c$ derived from the (1,1) line \\
\label{column_ratio}
\end{table}

In order to qualitatively investigate the origin of the observed chemical abundances variations we ran a grid of models using the chemical model UCL\_CHEM (\citealt{viti04}), a time and depth dependent gas-grain model extensively used to model a variety of astrophysical objects including low mass forming stars and prestellar cores (e.g. \citet{viti02}; \citet{roberts07}).
A detailed chemical modeling of the observed dense cores is not the aim of this work; instead we aim at determining whether the variation of the observed chemical abundances can be ascribed to a time evolution effect, in particular to the presence or absence of a protostar inside the dense core, or to differentiation in their physical characteristics, or a combination of both. Using UCL\_CHEM we follow the chemical and dynamical evolution of a collapsing core up to a chosen final density, which we treat as a free parameter. The initial gas density is at 100 cm$^{-3}$ and the gas is in atomic form (apart from a small fraction of hydrogen which is already in molecular form); the gas undergoes a free-fall collapse \citep{rawlings92} until the final densities are reached and it is at a temperature of 10 K. 
During this time atoms and molecules from the gas freeze on to the dust grains and they hydrogenate where possible. Note that the advantage of this approach is that the ice composition is not assumed but it is derived by a time dependent computation of the chemical evolution of the gas/dust interaction process. However, at 10 K, the ice composition does depend on i) the percentage of gas depleted on to the grains during the collapse, and this in turn depends on the density as well as on the sticking coefficient and other properties of the species and of the grains \citep{rawlings92};  ii) non thermal desorption processes (e.g. \citealt{roberts07}). UCL\_CHEM takes into consideration both thermal and non thermal desorption; the former is treated as in \citet{viti04} while the latter (which is due to local heating by H$_2$ formation on grains, cosmic rays and cosmic ray induced photons) as in \citet{roberts07}. 

The chemical network is adapted from the UMIST database and includes 221 species involved in 3194 gas-phase and grain reactions. 

We model two scenarios. The first scenario simulates the formation of a dense core (particle density higher than 5$\times$10$^4$ \cmtre) starting from a diffuse medium and following its chemical evolution in time. In the second scenario, after the formation of the dense core, a central protostar is included: we simulate the effect of the presence of an infrared source by subjecting the core to an increase in gas and dust temperature, up to 20 K  after the final density is reached. This increase in temperature leads to the sublimation of about 35\% of CO, N$_2$ and O$_2$ from the grains (see \citealt{collings04}; \citealt{viti04}); the chemical evolution of the gas is then followed up to 10$^7$ yrs. The first scenario represents, qualitatively, a starless dense core while the second scenario a protostellar core. We consider three final H$_2$ densities: 5$\times$10$^4$, 10$^5$ and 3$\times$10$^5$ \cmtre\, (considering that Lupus is a low mass star forming region) and a value of the product of the sticking probability and the grain cross section per unit volume (called FR), that defines the percentage of material on grains, between 0.1 and 0.8. These values correspond to a percentage of CO on grains at the end of the collapsing phase between 4\% and 55\%. In all the models the visual extinction is fixed at 10 mag, which is the average value found towards the cores \citep{chapman07}. We also run some models with A$_{\rm v}$=20 mag, which is the highest value found in the clouds, and we found, not surprisingly, that, at such high visual extinction, a difference of a factor of 2 does not influence the final abundance of the observed species. 
We investigated also the possibility of a collapse slower than free-fall by running models with a retarded collapse. We did not find large differences in the peak value of the chemical abundance of the investigated species apart from \hctren\, whose abundance is lower by about four to five orders of magnitude, and therefore it always remain well below the observed value. This is mainly due to the fact that in the case of a retarded collapse the gas is for a longer time at lower density (and hence at lower visual extinction); for species that easily photodissociate, such as \hctren, this implies a faster destruction. Of course the peak value of most molecules abundances is reached at longer times (t $\sim$ 3$\times$10$^7$ yrs) with respect to the free-fall model. We shall not consider models with a retarded collapse any further.

\begin{figure}
\includegraphics[width=6cm,angle=90]{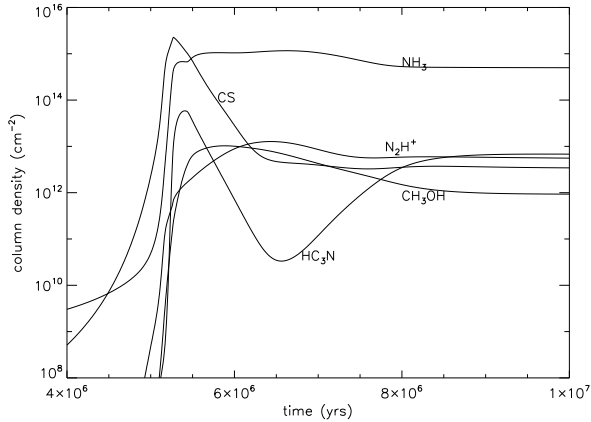}
\caption{Column densities {\it vs} time in the model for a starless dense cores with n(H$_2$)=5$\times$10$^4$ \cmtre\, and FR=0.2.}
\label{mod_pre}
\end{figure}

In all our models molecules start to form during the collapsing phase enhancing their abundance by several orders of magnitude. It is worth noting that the collapsing phase is the same in both scenarios, while they differentiate after the end of the collapse. 
A general trend that we see in all the models of starless cores, i.e. the first scenario, (an example is shown in Fig. \ref{mod_pre}) is that, once the final density is reached (at t=5.27$\times$10$^6$ yrs), the molecular abundances of the observed species evolve since they stabilize, changing by no more than a few hundredths with respect to the final value: the higher the depletion efficiency the shorter the time needed to reach the stable state. Another general trend is that the column density of \hctren\, has a double peak, the first maximum is reached soon after the final density is reached, the second is the final value. This double peak behavior was also found by other authors (e.g. \citealt{gwenlan00}). In Fig. \ref{mod_pre} we show the evolution of the column densities of the observed species in a starless core (i.e. in the first scenario) with a density of 5$\times$10$^4$ \cmtre\, and a FR=0.2, that corresponds to a percentage of CO on grains of 10\% at t=5.27$\times$10$^6$ yrs. At time around 9$\times$10$^6$ yrs the observed species stabilize their abundances and the predicted column densities of CS, \ndueh\, and \hctren\, fall in the range of the observed values while NH$_3$ and \chtreoh\, agree within a factor of 15. A good agreement with the observations is also found at early times, around 5.5$\times$10$^6$ yrs, for a slightly higher depletion efficiency (FR=0.4), however, also in this case, NH$_3$ is between one and two orders of magnitude higher then observed.

\begin{figure}
\includegraphics[width=6cm,angle=90]{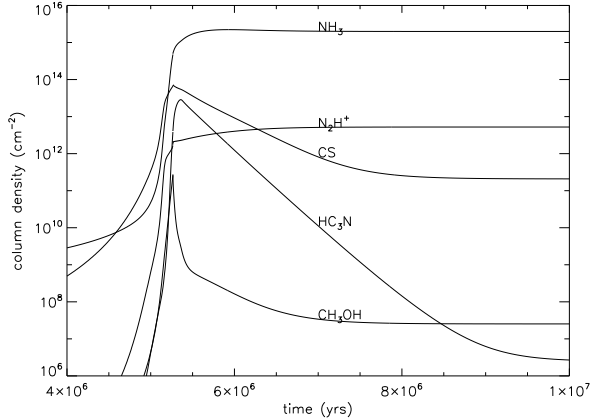}
\caption{Column densities {\it vs} time in the model for a protostellar dense cores with n(H$_2$)=5$\times$10$^4$ \cmtre\, and FR=0.5.}
\label{mod_pro}
\end{figure}

\begin{figure}
\includegraphics[width=6cm,angle=90]{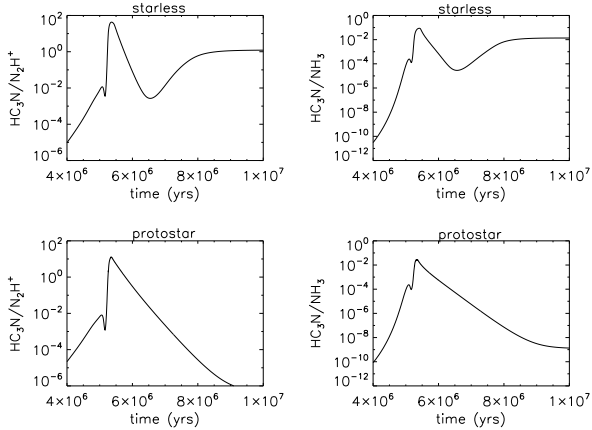}
\caption{Chemical abundances ratio of HC$_2$N/\ndueh\,(left) and HC$_3$N/NH$_3$ (right) {\it vs} time for a starless core (top, same model of Fig. \ref{mod_pre}) and a protostar (bottom, same model of Fig. \ref{mod_pro}).}
\label{mod_ratio}
\end{figure}

In Fig. \ref{mod_pro} we show the column densities of the observed species vs time in the best fit model of the second scenario, i.e. where a young protostar is present in the core. The general behaviour of CS, NH$_3$ and \ndueh is similar to the previous scenario while \hctren\, and CH$_3$OH behave differently. In particular \hctren\, does not show the second peak and CH$_3$OH decreases quickly. The reason why methanol decreases so quickly in the second scenario is that the partial release of CO leads to an increase of CH that seems to effectively react with CH3OH.
In the model with gas density of 5$\times$10$^4$ \cmtre\, and FR=0.5, that corresponds to a percentage of CO on grains of 20\% at t=5.27$\times$10$^6$ yrs, (Fig. \ref{mod_pro}) the best agreement with the observation is at t=5.3$\times$10$^6$ yrs when all the column densities agree with the observed values within one order of magnitude apart from \chtreoh\, that is 2 orders of magnitude lower. This model reproduces also the observed values of HC$_3$N/\ndueh\, ratio even if it predicts a HC$_3$N/NH$_3$ ratio about a factor of 10 -- 100 lower than observed (see Fig. \ref{mod_ratio}). In particular the higher observed value of HC$_3$N/\ndueh\, (around 10) is reached as soon as the collapse is stopped, at t $\sim$ 5.3$\times$10$^6$ yrs, while the lower observed value (around 1) is reached at t $\sim$ 6$\times$10$^6$ yrs.

In general, models with gas density of 5$\times$10$^4$ \cmtre\, and CO depletion between 15\% and 30\% predict column densities, at t$\sim$5.3$\times$10$^6$ yrs, that agree within at maximum a couple of orders of magnitude with the observations, at later time the methanol is too low. Higher depletion predicts HC$_3$N/\ndueh $\leq$1 and a maximum of the \hctren\, column density several orders of magnitude lower than observed. On the other hand, models with CO depletion lower than 15\% reproduce the observed \hctren\, column density but a HC$_3$N/\ndueh\, ratio up to 100  and a CS column density 3 orders of magnitude higher than observed.
A general trend that we find in all our models is that the maximum column densities of NH$_3$ and CH$_3$OH change by no more than a factor of 3 and \ndueh change by no more than a factor of 10 within the explored range of model parameters. On the other hand, CS and \hctren\, are quite sensitive to the depletion efficiency, decreasing their maximum abundance by several orders of magnitude as the depletion increases.

In all our models we see that once the dense core is formed, \hctren\, starts to decrease while \ndueh\, and NH$_3$ slightly increase. In other words HC$_3$N/\ndueh\, and HC$_3$N/NH$_3$ decrease after the collapse phase is halted (see Fig. \ref{mod_ratio}) while NH$_3$/\ndueh\, increases, thus suggesting that the abundances of these species may be indicators of the chemical evolution of dense cores. In fact, in the starless scenario \hctren, after the decreasing phase, increases again till it reaches a constant value where the HC$_3$N/\ndueh\, ratio is around 1 and HC$_3$N/NH$_3$ around 0.01.
In summary, higher values of HC$_3$N/\ndueh\, and HC$_3$N/NH$_3$ are expected in dense cores both starless or associated with a very young protostar; more evolved protostars (probably late Class 0 or Class I protostars) may have lower values of both HC$_3$N/\ndueh\, and HC$_3$N/NH$_3$.

Considering the observed value of the HC$_3$N/\ndueh\, ratio we can classify Lup1 C1, Lup1 C2, Lup1 C3, Lup1 C5, Lup1 C8 and Lup3 C5 as very young protostars or starless dense cores and Lup1 C4, Lup1 C6, Lup1 C7, Lup3 C1, Lup3 C2, Lup3 C3 and Lup3 C4 as more evolved protostars. The classification of the two cores detected in Lupus 4 is difficult since only \hctren\, and NH$_3$ have been observed in this cloud. However, from the abundance ratio of the two species, we can speculate the Lup4 C2 is younger than Lup4 C1.

\section{On the evolutionary status of the clouds}

We can compare our classification with the results of the Spitzer survey carried out in the c2d Legacy Program (\citealt{chapman07}; \citealt{merin08} ). As commonly known the Spitzer surveys, covering the 3--160 \um\, spectral range, are efficient in discovering and classifying YSOs in the Class I, II and III evolutionary stage. On the contrary, the sources that are in a previous evolutionary stage are not detected in the Spitzer surveys since the SED of these younger sources peaks at longer wavelengths then the Spitzer bands. 

Our evolutionary classification of the dense cores identified in the Mopra maps is supported by the analysis of the Spitzer survey. Indeed only two of our millimetre cores are also detected by Spitzer i.e. Lup1 C4, which is associated with source 10 and classified as Flat  by \citet{merin08} (the same source was previously classified as Class I by \citet{chapman07}) and Lup3 C3 which is associated with source 87 by \citet{merin08} and classified as Class I. It is worth noting that looking at the HC$_3$N/\ndueh column density ratio, Lup1 C4 and Lup3 C3 seem to be two of the most evolved sources of our sample.
Another comparison can be made with the H$^{13}$CO$^+$ and 1.2 mm maps of Lupus 3 by \cite{tachihara07}. By means of a SED analysis, core Lup3 C3 has been classified by \cite{tachihara07} as a Class 0 object while cores Lup3 C2, C4 and C5 have been classified as pre-stellar cores. These classifications agree with our data since Lup3 C3 is the most prominent core in NH$_3$ and Lup3 C5 is the most prominent core in HC$_3$N; Lup3 C2 and Lup3 C4 seem to be at an intermediate stage between the former two since they are bright in NH$_3$ and \ndueh but also weakly emit in \hctren.

Our study shows that Lupus 1 is the cloud richest in millimetre dense cores, most of those  likely being protostellar cores. On the other hand, the Spitzer surveys have shown that Lupus 3 is richer in YSOs with respect to Lupus 1 and 4; indeed 69, 13 and 12 YSOs have been detected in Lupus 3, 1 and 4 respectively \citep{merin08}. 
The ratio between YSOs candidates, i.e. Class I, II and III objects detected by Spitzer and the millimetre dense cores i.e. prestellar and young protostellar cores detected in our maps is 13.8, 6.0 and 1.6 in Lupus 3, 4 and 1 respectively. All these evidences point towards a picture where in Lupus 3 the bulk of the star formation activity has already passed and only a moderate number of stars are still forming in the eastern part. On the contrary, in Lupus 1 star formation is on-going, with several dense cores still in the pre--stellar phase. Lupus 4 is at an intermediate stage, with a smaller number of individual objects.

Age differences between the stars in the Lupus clouds have been already noticed by \cite{hughes94} analyzing the histogram of the ages distribution of the stellar population, from which it appears that Lupus 3 and 4 are more evolved than Lupus 1. \cite{tachihara96} explained the relative youth of Lupus 1 as well as Lupus 2 as resulting from a recent passage of a shock associated with the expanding Upper Scorpius shell across the cloud.

\section{Conclusions}

The molecular clouds Lupus 1, 3 and 4 were mapped with the Mopra telescope at 3 and 12 mm. Emission lines from high density molecular tracers were detected, i.e.  NH$_3$ (1,1) at 23.694 GHz, NH$_3$ (2,2) at 23.722 GHz, HC$_3$N (3$_4$-2$_3$) at 27.294 GHz, CS (2-1) at 97.981 GHz, CH$_3$OH (2$_0$-1$_0$)A$^+$ at 96.741 GHz, CH$_3$OH (2$_{-1}$-1$_{-1}$)E at 96.739 GHz, N$_2$H$^+$ (1-0) at 93.174 GHz and HC$_3$N (10-9) at 90.979 GHz. In the CS channel maps we found velocity gradients of more of 1 \kms\, across both Lupus 1 and 3. Moreover, towards the source Lup1 C6 and in the southern part of Lupus 3 (Lup3 C2 and Lup3 C4) two different velocity components are present along the same line of sight. 

A total of 15 high density gas cores were detected in the three clouds and the column density of the observed species towards each core was derived. In the five cores where both NH$_3$ (1,1) and (2,2) lines were detected we derived a kinetic temperature between 12 and 13 K, a value typical of the cold dense cores associated with protostars. The cores are roughly spherical at our resolution, with sizes ranging from 1\arcmin\, to 2\arcmin. The cloud richest in high density cores is Lupus 1 where 8 cores have been detected, 5 cores were detected in Lupus 5 and only 2 in Lupus 4. 

The intensity of the lines of the three species \hctren, NH$_3$ and \ndueh changes significantly between the various cores: cores that are brighter \hctren\, are fainter or undetected in NH$_3$ and \ndueh and vice versa.
This behavior was quantified by considering the ratio between the column densities that reflects the ratio of the chemical abundance of the species if they come from the same region. We found that the two ratios HC$_3$N/\ndueh and HC$_3$N/NH$_3$ change by one order of magnitude between the cores. We use the time dependent UCL\_CHEM chemical code to qualitatively reproduce two scenarios: i) a starless dense core and ii) a core containing a protostar. 
Models with gas density of 5$\times$10$^4$ \cmtre\, and CO depletion between 15\% and 30\% predict column densities at t$\sim$5.3$\times$10$^6$ yrs that agree within at maximum a couple of orders of magnitude with the observations. In starless core models a good agreement can be also find at later time, t$\geqslant$9$\times$10$^6$ yrs, for gas density of 5$\times$10$^4$ \cmtre\, and CO depletion percentage around 10\%. In all our models we found that, after the collapsing phase, the HC$_3$N/\ndueh\,and HC$_3$N/NH$_3$ ratios decrease with time, thus indicating that the abundances of these species may be indicators of the chemical evolution of dense cores. On this base we classified 5 out of 8 cores in Lupus 1 and 1 out of 5 cores in Lupus 3 as prestellar cores or very young protostars. We compared our millimetre maps with the Spitzer survey that detected more evolved YSOs and we found that only two cores were also detected by Spitzer, thus confirming the youth of the cores observed with Mopra. The ratio between YSOs candidates, i.e. Class I, II and III objects detected by Spitzer and the millimetre dense cores i.e. prestellar and young protostellar cores detected in our maps is 13.8, 6.0 and 1.6 in Lupus 3, 4 and 1 respectively. 

We conclude that in Lupus 3 the bulk of the star formation activity has already passed and only a moderate number of stars are still forming in the eastern part. On the contrary, in Lupus 1 star formation is on-going with several dense cores still in the pre--stellar phase. Lupus 4 is at an intermediate stage, with a smaller number of individual objects.

\section*{Acknowledgments}
We thank G. Busquet for the useful comparison with her chemical models.
The Mopra Telescope is part of the Australia Telescope and is funded by the Commonwealth of Australia for operation as National Facility managed by CSIRO. The University of New South
Wales Mopra Spectrometer Digital Filter Bank used for the observations with the Mopra Telescope was provided with support from the Australian Research Council, together with the University
of New South Wales, University of Sydney and Monash University.

\appendix

\section[]{Column density determination}

We used different methods for the calculation of the column density. 
In fact, for the \ndueh (1-0), NH$_3$ (1,1) and \hctren (3-2) transitions the optical depth and the excitation temperature have been derived from the hyperfine fitting whilst for the other lines these two crucial parameters cannot be derived, therefore we must assume the optical thin and LTE approximation. 

For \ndueh (1-0) and \hctren (3-2) the column density is given by the following formula (from \cite{caselli02b}) valid for optically thick transitions
\begin{equation}
N =  \frac{8 \pi^{3/2}}{1.6651} \frac{\nu^3}{c^3 g_u A_{ul}} ~
\frac{e^{\left(\frac{E_l}{kT_{ex}}\right)}}{1-e^{\left(\frac{-h \nu}{kT_{ex}}\right)}} ~Q ~\tau~ \Delta v ~~ cm^{-2}
\label{caselli}
\end{equation}
where $\Delta v$ is the line width, $\nu$ is frequency of the observed transition, $A_{ul}$ is the Einstein coefficient, $g_u$ are the statistical weight of the upper level, $\tau$ is the optical depth, $T_{ex}$ is the excitation temperature, Q is the partition function, $E_l$ is the energy of the lower level.
In particular, for the NH$_3$ (1,1) line we used an approximated formula derived by \cite{bachiller87}, also valid for optically thin lines
\begin{equation}
N({\rm NH}_3 (1,1))= 2.784\times10^{13} \tau T_{ex} \Delta v ~~ cm^{-2}
\label{bachiller}
\end{equation}

For \hctren (10-9), CS (2-1) and the two methanol lines we  assume the optical thin and LTE approximation and the column density is given by the following formula

\begin{equation}
N=\frac{8\times10^5 \pi k \nu^2 e^{\left(\frac{E_u}{kT_{rot}}\right)} }{hc^3g_{up}A_{ud}} ~ Q(Trot) ~\int{T_{mb}~ dv}
\end{equation}

as $T_{rot}$ we assume the temperature derived from NH$_3$ or, for the cores where there is not this estimate, assuming the typical value derived in dense cores, T=10 K \citep{tafalla02}. The molecular data are taken from the JPL catalog\footnote{http://spec.jpl.nasa.gov/home.html}, apart the Einstein coefficients that are taken from the CDMS catalog\footnote{http://www.astro.uni-koeln.de/cdms}.

The partition function of the linear species is
\begin{equation}
 Q(T_{rot})=\sigma\frac{kT_{rot}}{hB}
\end{equation}
The partition function of methanol, that is and asymmetric tops species, has been derived from the linear interpolation of the values given in the JPL catalog.

\section[]{Kinetic temperature determination}
From the two lines NH$_3$ (1,1) and (2,2) we derived the rotational temperature following the \citet{ungerechts86} approach by using the following formula

\begin{equation}
T_{rot}=\frac{-41.5}{ln\{- \frac{0.282}{\tau_{(1,1,m)}} ln[1-\frac{Ta_{(2,2,m)}}{Ta_{(1,1,m)}} (1-e^{-\tau_{(1,1,m)}})]\}}
\label{unger86}
\end{equation}
where $\tau_{(1,1,m)}$ is the optical depth of the main component of the (1,1) transition (derived by the hyperfine fitting as described in Sect. 2.1 and reported in Table \ref{hyperfine}) and Ta$_{(2,2,m)}$ and Ta$_{(1,1,m)}$ are the antenna temperatures of the main component of the (2,2) and (1,1) transitions, respectively.

From the rotational temperature we derived the kinetic temperature applying a numeric approximation \citep{tafalla04} valid in the low temperature regime (5$\leq$T$_{kin}\leq$20 K)
\begin{equation}
T_{kin}= \frac{T_{rot}}{1-\frac{T_{rot}}{42} ln(1+1.1 exp(-16/T_{rot}))}
\label{tafalla}
\end{equation}

\bsp

\label{lastpage}

\end{document}